\documentclass{sn-jnl}

\usepackage{natbib}
\usepackage{booktabs}
\usepackage{graphicx}%
\usepackage{multirow}%
\usepackage{amsmath,amssymb,amsfonts}%
\usepackage{amsthm}%
\usepackage{mathrsfs}%
\usepackage[title]{appendix}%
\usepackage{xcolor}%
\usepackage{textcomp}%
\usepackage{manyfoot}%
\usepackage{multicol}
\usepackage{algorithm}%
\usepackage{algorithmicx}%
\usepackage{algpseudocode}%
\usepackage{listings}%

\usepackage{hyperref}
\hypersetup{
citecolor= cyan, 
    colorlinks=true,
    linkcolor=cyan, 
    filecolor=magenta,      
    urlcolor=magenta,
    pdftitle={Overleaf Example},
    pdfpagemode=FullScreen,
    }
\newcommand{\add}[1]{\textcolor{blue}{#1}}

\usepackage{ulem}

\begin{document}

\title{Spatio-temporal point process modelling of fires in Sicily exploring human and environmental factors}


\author*[1]{\fnm{Nicoletta} \sur{D'Angelo}}\email{nicoletta.dangelo@unipa.it}

\author[1,2]{\fnm{Alessandro} \sur{Albano}}\email{alessandro.albano@unipa.it}

\author[3]{\fnm{Andrea} \sur{Gilardi}}\email{andrea.gilardi@polimi.it}

\author[1]{\fnm{Giada} \sur{Adelfio}}\email{giada.adelfio@unipa.it}

\affil*[1]{\orgdiv{Department of Economics, Business and Statistics}, \orgname{University of Palermo}, \orgaddress{\city{Palermo},  \country{Italy}}}

\affil[2]{\orgdiv{Sustainable Mobility Center (Centro Nazionale per la Mobilità Sostenibile—CNMS)}}

\affil[3]{\orgdiv{MOX - Department of Mathematics}, \orgname{Politecnico di Milano}, \orgaddress{\city{Milano},  \country{Italy}}}


\abstract{In 2023, Sicily faced an escalating issue of uncontrolled fires, necessitating a thorough investigation into their spatio-temporal dynamics. Our study addresses this concern through point process theory. Each wildfire is treated as a unique point in both space and time, allowing us to assess the influence of environmental and anthropogenic factors by fitting a spatio-temporal separable Poisson point process model, with a particular focus on the role of land usage. First, a spatial log-linear Poisson model is applied to investigate the influence of land use types on wildfire distribution, controlling for other environmental covariates. The results highlight the significant effect of human activities, altitude, and slope on spatial fire occurrence. Then, a Generalized Additive Model with Poisson-distributed response further explores the temporal dynamics of wildfire occurrences, confirming their dependence on various environmental variables, including the maximum daily temperature, wind speed, surface pressure, and total precipitation.}


\keywords{Fires, Land usage, Point processes, Spatial analysis, Intensity estimation.}



\maketitle

\section{Introduction}

In recent years, Sicily has experienced an alarming increase in uncontrolled wildfires, establishing it as the Italian region with the highest frequency of fire events and, consequently, the largest burned area. The occurrence of these fires is primarily associated with ignition sources, forest fuels, and environmental conditions \citep{ganteaume2013review,hantson2015global,ricotta2014modeling}. Ignition sources are typically categorized into natural causes, such as lightning and geological factors, and human causes, both accidental and intentional \citep{aldersley2011global,rodrigues2014insight}. Human-driven factors, particularly arson (but also, unintentional wildfires), often ignited for purposes like creating new pasture resources or burning stubble, stand out as the primary contributors to wildfires in Sicily, especially in areas where vegetation interfaces with urban structures \citep{ferrara2019background}.

Wildfires pose a significant threat to ecosystems, human settlements, and economic activities, demanding a comprehensive understanding of their spatio-temporal dynamics for effective mitigation and management. The island of Sicily, nestled in the heart of the Mediterranean Sea, has a rich history and diverse landscapes, making it vulnerable to the escalating impacts of climate change, including an increased frequency and intensity of wildfires. In the year 2023, Sicily faced a pronounced wildfire season, underscoring the urgent need for advanced analytical methodologies to unravel the underlying patterns and drivers of these destructive events.

Traditional approaches to wildfire analysis often rely on descriptive statistics and basic spatial visualization, providing limited insights into the effect of environmental, climatic, and anthropogenic factors that contribute on wildfire occurrences. This paper advocates for the application of point process methodology to explore the spatial and temporal dynamics of Sicilian wildfires in 2023.

Point process methodology, rooted in statistical theory, enables the modeling of events occurring in space and time, making it an ideal tool for in-depth analysis of wildfire patterns. By treating each wildfire occurrence as a point in a spatial and temporal domain, we aim to discern potential predictors that may contribute to the ignition of wildfires across the Sicilian landscape.

Moreover, traditional analyses often fall short of capturing the relationships between environmental, climatic, and anthropogenic factors that contribute to wildfires. By exploiting spatio-temporal models, we aim to assess which variables influence the probability of fire occurrence. Including covariates such as land usage, altitude, slope, temperature, precipitation, surface pressure, and wind speed provides a comprehensive framework to understand the underlying mechanisms influencing wildfire patterns.

As noted by \cite{butsic2015land}, fires and land usage are intrinsically connected, but research investigating this dynamic is limited.
Specifically, the influence of land use on fire ignition is a well-documented phenomenon. Previous papers have shown that land use, particularly the presence of roads, can significantly impact the ignition of fires \citep{ricotta2018assessing}, with the influence of roads being much stronger in less dense land cover areas.
As human activities increasingly encroach upon natural landscapes, understanding the impact of land usage becomes crucial for effective wildfire management. Through the proposed modelling, we seek to quantify the relationship between different land use types and the probability of fire occurrence, finding areas where human activities may be making the Sicilian landscape worse.

The proposed spatio-temporal separable Poisson point process model seeks to determine the specific contributions of land usage categories on fire occurrences while controlling for other environmental covariates. As a matter of fact, understanding the influence of land usage on wildfire probability is crucial for designing targeted interventions, land-use planning, and developing resilient ecosystems in the face of evolving environmental challenges.
We first fit a spatial log-linear Poisson model in order to investigate the influence of land use types on wildfire occurrence, controlling for other environmental covariates. The results highlight the significant effect of human activities, altitude, and slope on spatial fire occurrence. Secondly, a Generalized Additive Model with Poisson-distributed response is fitted to model the temporal fire occurrences, confirming their dependence on various environmental variables, including the maximum daily temperature, wind speed, surface pressure, and total precipitation.
This study seeks to contribute not only to the scientific understanding of wildfire dynamics but also to the practical tools necessary for mitigating their impacts on communities and ecosystems.

All the analyses are carried out through the \texttt{R} statistical software \citep{R}, making particular use of the packages \texttt{stopp} \citep{stopp}, \texttt{sf} \citep{Rsf}, and \texttt{stars} \citep{Rstars}.

The structure of the paper is as follows.  Section \ref{sec:analysis} presents the data. Section \ref{sec:met} provides an overview of spatio-temporal point processes and the model employed in the paper, while Section \ref{sec:int} illustrates the actual model fitting through the spatial and temporal intensity estimation. Section \ref{sec:concls} is devoted to the discussion and conclusions. 

\section{Data}\label{sec:analysis}

This section focuses on the description of the data employed in the paper, namely the fire data, which will represent our spatio-temporal point pattern under analysis, and some spatial and/or temporal variables which will serve as covariates, assumed to influence the overall occurrence of points, as typical in point process analysis.

\subsection{Fires point pattern}\label{sec:data}
To conduct our analysis, we rely on data obtained from the Fire Information for Resource Management System (FIRMS) platform, accessible for download at the following URL: \url{https://firms.modaps.eosdis.nasa.gov/download/}. This source provides near real-time active fire locations to natural resource managers.

The variables considered in our study are:

\begin{itemize}
    \item \texttt{Latitude}: Center of 1 km fire pixel, but not necessarily the actual location of the fire as one or more fires can be detected within the 1 km pixel.

    \item \texttt{Longitude}: Center of 1 km fire pixel, but not necessarily the actual location of the fire as one or more fires can be detected within the 1 km pixel.

    \item \texttt{Acq\_Date} (Acquisition Date): Date of MODIS acquisition.

    \item \texttt{Acq\_Time} (Acquisition Time): Time of acquisition/overpass of the satellite (in UTC).

\end{itemize}

In particular, \texttt{Longitude} and \texttt{Latitude} will be used as spatial coordinates of our point pattern, while \texttt{Acq\_Date} and \texttt{Acq\_Time}  will be combined in a unique variable serving as the time occurrence of the fire, with the smallest detail as the hour \add{of} occurrence within a day.

\begin{figure}[!h]
\centering
\includegraphics[width=.75\textwidth, trim = {0 1.5cm 0 2cm}, clip]{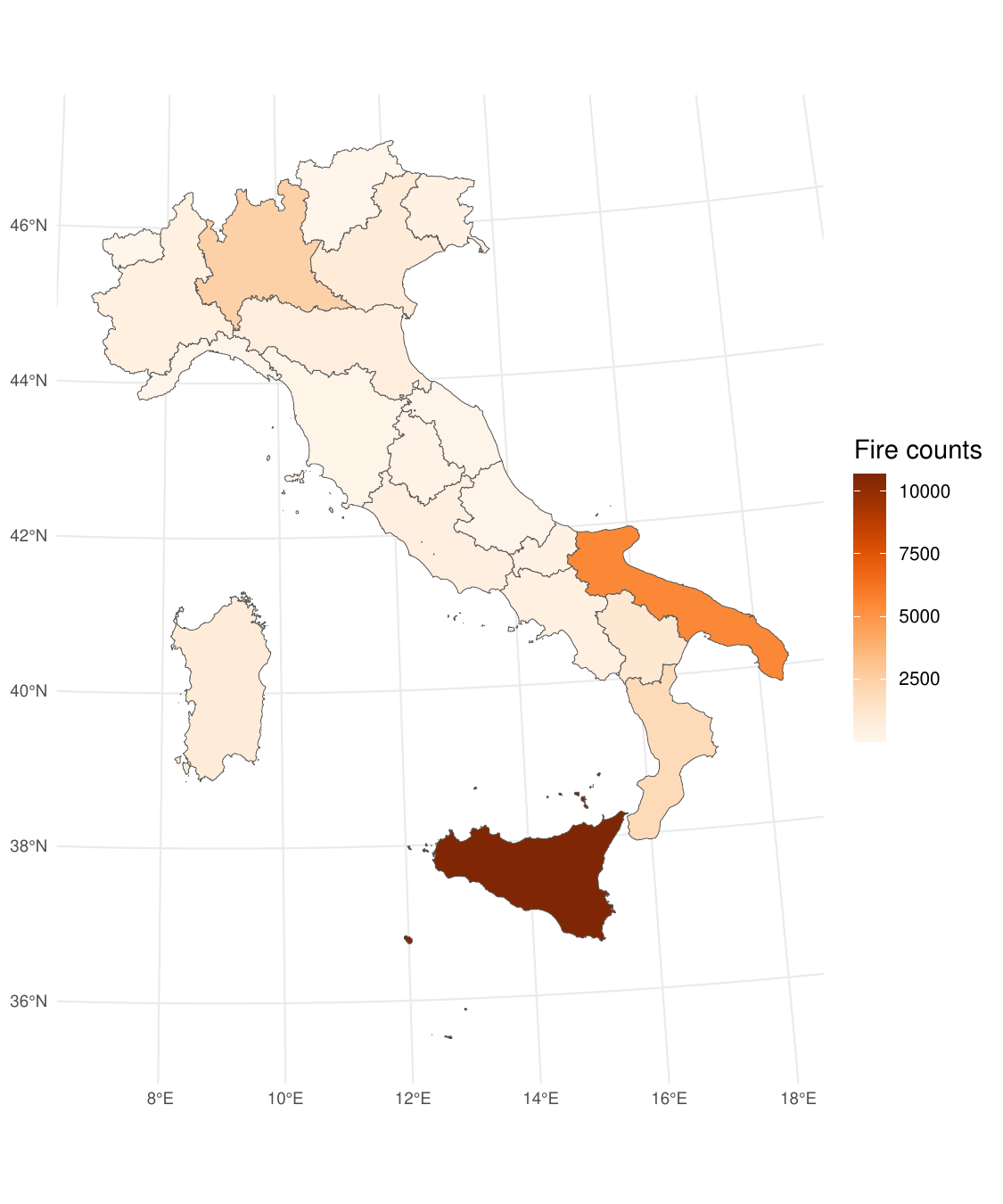}	
\caption{Choropleth map displaying the counts of the wildfires that occurred in each Italian region during 2023.}
 \label{fig:0}
\end{figure}

\begin{figure}[!h]
  \centering
   \includegraphics[width=.75\textwidth]{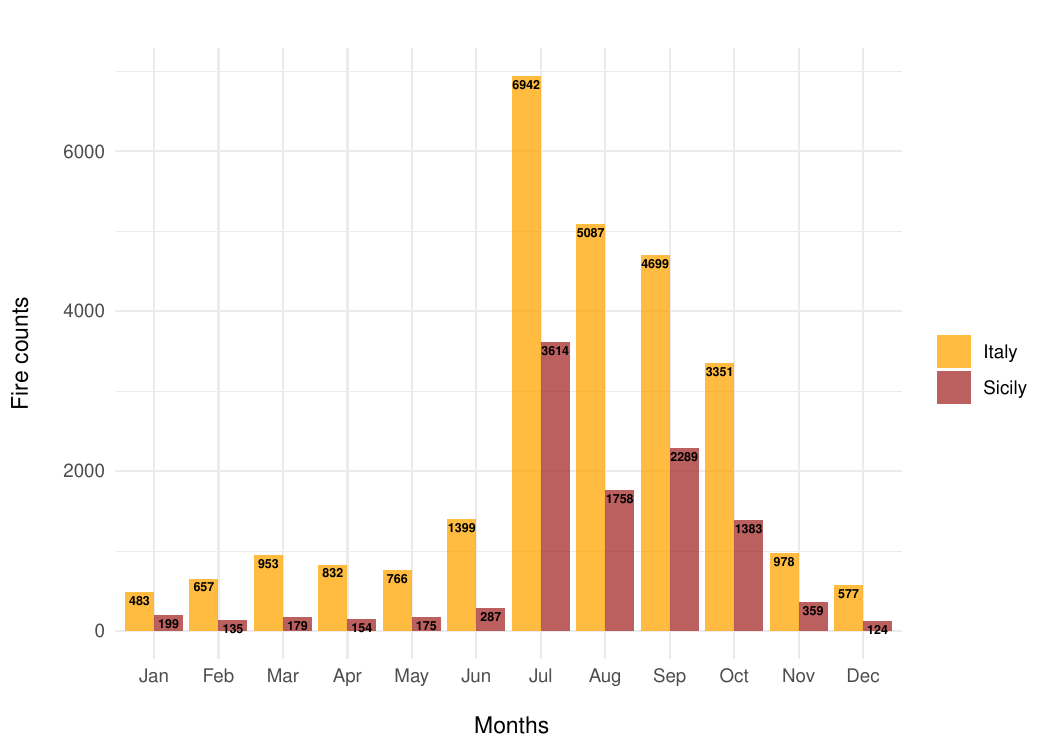}	
	\caption{Barplot comparing the number of fires that occurred in the whole country (yellow bars) and just in the Sicily region (red bars).}
	\label{fig:5b}
\end{figure}

\begin{figure}[!h]
\centering
\includegraphics[width=\textwidth, trim = {1.25cm 1.75cm 0 1.6cm}, clip]{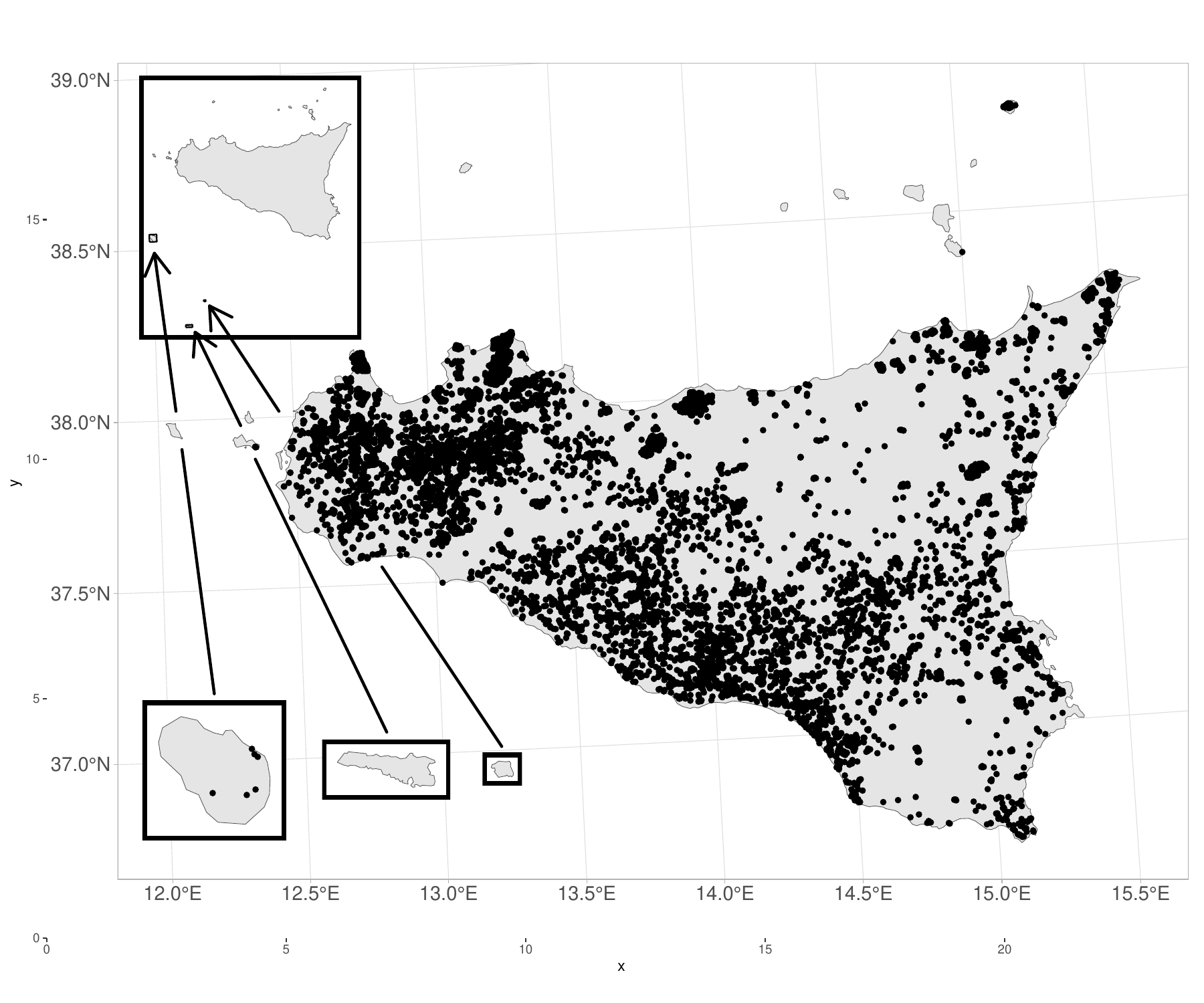}	
\caption{Spatial distribution of fires recorded in the Sicily region during 2023. The inset map reported in the top-left displays the whole observation window which includes the mainland and three smaller islands (namely Pantelleria, Lampedusa, and Linosa, respectively), located in the far South and reported in the bottom-left as inset maps.}
\label{fig:000}
\end{figure}

Focusing exclusively on the spatial distribution, Figure \ref{fig:0} presents fire counts in Italy throughout the year 2023. 
In particular, it shows the counts by region of the 26,724 fires that occurred in Italy in 2023. 
Notably, the graphical representation emphasizes a pronounced concentration of fires in the Sicilian territory, with the second-highest incidence observed in the Southern region of Puglia (\textit{Apulia}).
Following this, Figure \ref{fig:5b} compares fire counts that occurred in Italy and Sicily in 2023.
It is evident that the summer months, from July to October, stand out as critical periods with the highest number of fires. Specifically, July emerges as the most challenging month, recording a total of 8,842 wildfires in Italy, of which 3,814 occurred exclusively in Sicily. Finally, Figure \ref{fig:000} shows the spatial distribution of Sicilian fires in 2023. The inset map (which is reported in the top-left) displays the whole observation window, which includes the mainland of the region and three smaller islands (namely Pantelleria, Lampedusa, and Linosa, respectively), located in the far South and reported in the bottom-left as inset maps. The figure notably indicates clustering behaviour, particularly in the western parts of the mainland.

\subsection{Land use}

The main focus of our work is the analysis of the fire occurrences depending on the land usage since, as noted later, this type of information can be used as a spatial covariate for our purposes. The land use data come from the CORINE Land Cover (CLC) dataset\footnote{URL: \url{https://land.copernicus.eu/en/cart-downloads}; Data downloaded during \add{2023-11}}, which is a comprehensive land cover and land-use database created by the European Environment Agency (EEA) to facilitate environmental monitoring and assessment across Europe. It is usually employed to provide insights into the changing landscapes of Europe, aiding in several projects such as environmental management and spatial planning, in addition to wildfire analysis as exemplified in this paper. Combining satellite imagery (obtained from Sentinel-2 and Landsat-8 projects) and ground-based information, CLC classifies land cover into a hierarchical raster system of over 40 classes, including urban areas, forests, water bodies, and agricultural lands. The detailed classification according to CLC comes in Table \ref{tab:land_use}, whereas the product documentation and the nomenclature guidelines can be browsed on the project's website\footnote{URL: \url{https://land.copernicus.eu/content/corine-land-cover-nomenclature-guidelines/html/}. Last access: Feb. 2024.}. 

\begin{table}[!h]
  \centering
    \caption{Land Use Classification according to Corine Land Cover Legend.}
 \begin{tabular}{l|l|l}
    \toprule
 \textbf{Level 1} & \textbf{Level 2}& \textbf{Level 3} \\ \midrule
    Artificial surfaces &Urban fabric& Continuous urban fabric \\ 
     && Discontinuous urban fabric \\ 
      & Industrial, commercial & Industrial or commercial units \\ 
      &and transport units& Road and rail networks and associated land \\ 
     && Port areas \\ 
      && Airports \\ 
      & Mine, dump& Mineral extraction sites \\ 
     & and construction sites& Dump sites \\ 
     && Construction sites \\ 
      &Artificial, non-agricultural & Green urban areas \\ 
     &vegetated areas& Sport and leisure facilities \\ \midrule
    Agricultural areas & Arable Land& Non-irrigated arable land \\ 
      && Permanently irrigated land \\ 
      & Permanent crops& Vineyards \\ 
      && Fruit trees and berry plantations \\ 
      && Olive groves \\ 
     & Heterogeneous & Annual crops associated with permanent crops \\ 
      &agricultural areas& Complex cultivation patterns \\ 
      && Land principally occupied by agriculture \\
      && with significant areas of natural vegetation \\ \midrule
    Forest and semi-natural areas &Forests& Broad-leaved forest \\ 
      && Coniferous forest \\ 
      && Mixed forest \\ 
      &Scrub and/or herbaceous & Natural grasslands \\ 
      &vegetation associations& Moors and heathland \\ 
     && Sclerophyllous vegetation \\ 
     && Transitional woodland-shrub \\ 
      &Open spaces with little& Beaches dunes sands \\ 
      & or no vegetation& Bare rocks \\ 
      && Sparsely vegetated areas \\ 
     && Burnt areas \\ \midrule
    Wetlands  &Inland wetlands& Inland marshes \\ 
      &Marine wetlands& Salt marshes \\ 
      && Salines \\ \midrule
    Water bodies &Inland waters& Water courses \\ 
      && Water bodies \\ 
      &Marine waters& Coastal lagoons \\ 
      && Sea and ocean \\ \bottomrule
  \end{tabular}
  \label{tab:land_use}
\end{table}

\begin{figure}[!h]
\centering
   	\includegraphics[width=\textwidth, trim = {1.25cm 3.5cm 0.25cm 3.5cm}, clip]{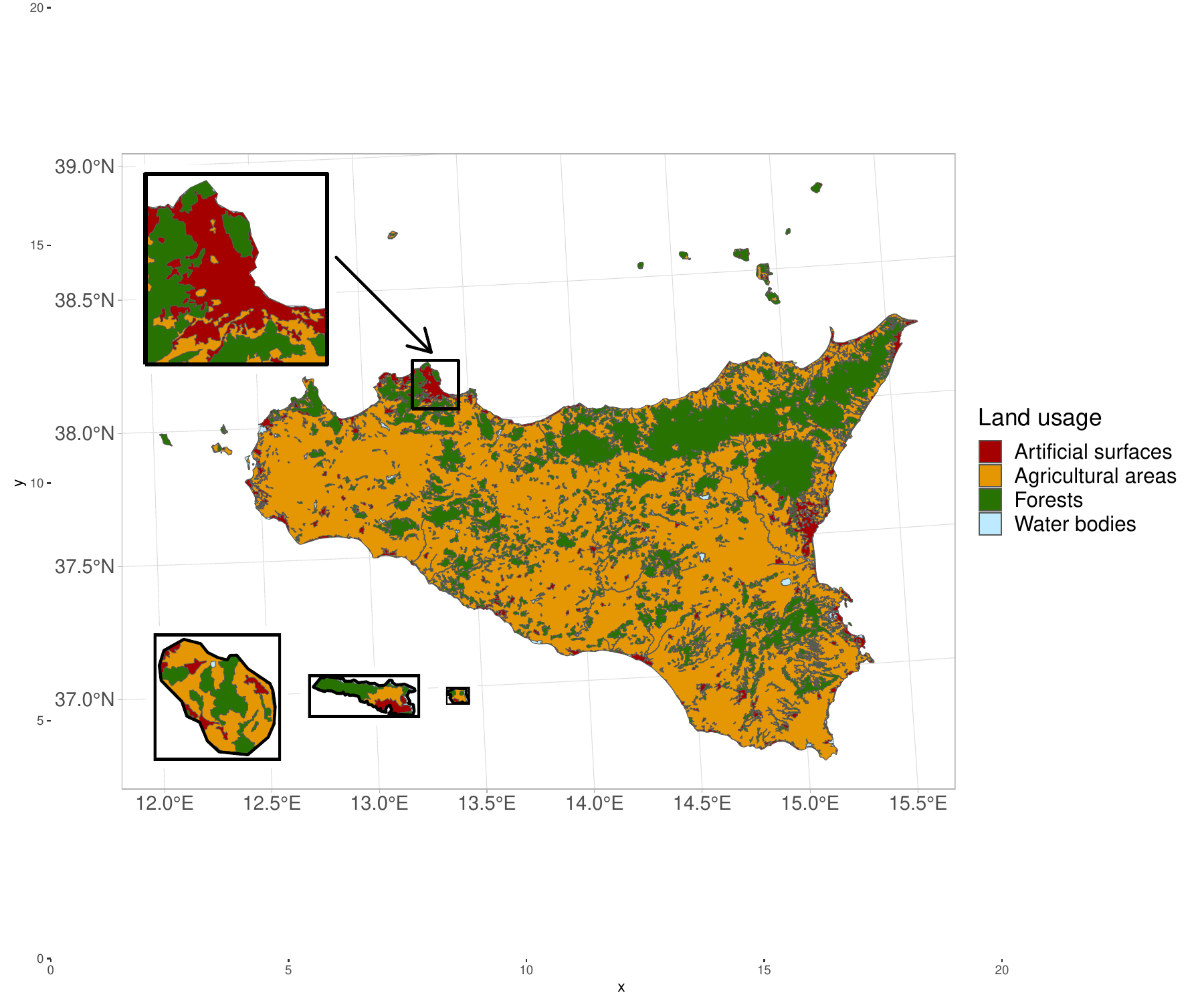}	
	\caption{Land usage in Sicily as reported in the CORINE Land Cover dataset. The inset map included in the top-left zooms around the area of Palermo (which in the capital of the region), whereas the three inset maps displayed in the bottom-left showcase the land usage for Pantelleria, Lampedusa, and Linosa.}
	\label{fig:11}
\end{figure}

Figure~\ref{fig:11} depicts the downloaded data on land usage categorized by level 1 of the CLC Legend after merging the 4th and 5th categories, i.e. wetlands and water bodies (``waterbodies" from now on). In particular, we represent the macro land usage classification, whose categories are: \begin{itemize}
    \item Artificial surfaces
    \item Agricultural areas
    \item Forest and semi-natural areas
    \item Water bodies
\end{itemize}

As shown in Table \ref{tab:land_use2}, the majority of the Sicilian territory is constituted by agricultural areas ($68\%$, mainly in the souther parts of the region), followed by forests and semi-natural areas ($26\%$, typically located in proximity of some mountain chains in the North and North-East areas). Then, artificial surfaces and water bodies represent the smallest portion ($5\%$ and $1\%$, respectively). 

\begin{table}[!h]
\centering
\caption{Area percentage, number of fires, and number of fires over the area percentage of the macro land usage.}
\begin{tabular}{lrrr}
  \toprule
 & \textbf{Area perc.} & \textbf{Points} & \textbf{Points perc.} \\ 
  \midrule
 \textbf{Artificial surfaces}  &0.050 & 858 & 0.081\\ 
   \textbf{Agricultural areas}  & 0.683 & 6529 & 0.613\\ 
   \textbf{Forest and semi-natural areas} &  0.259 & 3224 & 0.303  \\ 
   \textbf{Water bodies} & 0.007 & 45 & 0.004\\ 
   \bottomrule
\end{tabular}
\label{tab:land_use2}
\end{table}

\subsection{Environmental covariates}

The other covariates considered in this work are now introduced, detailing the sources from where we downloaded the raw data and explaining the pre-processing operations adopted to convert them into an usable format. 

\begin{figure}
    \centering
    \includegraphics[width = 0.4\linewidth]{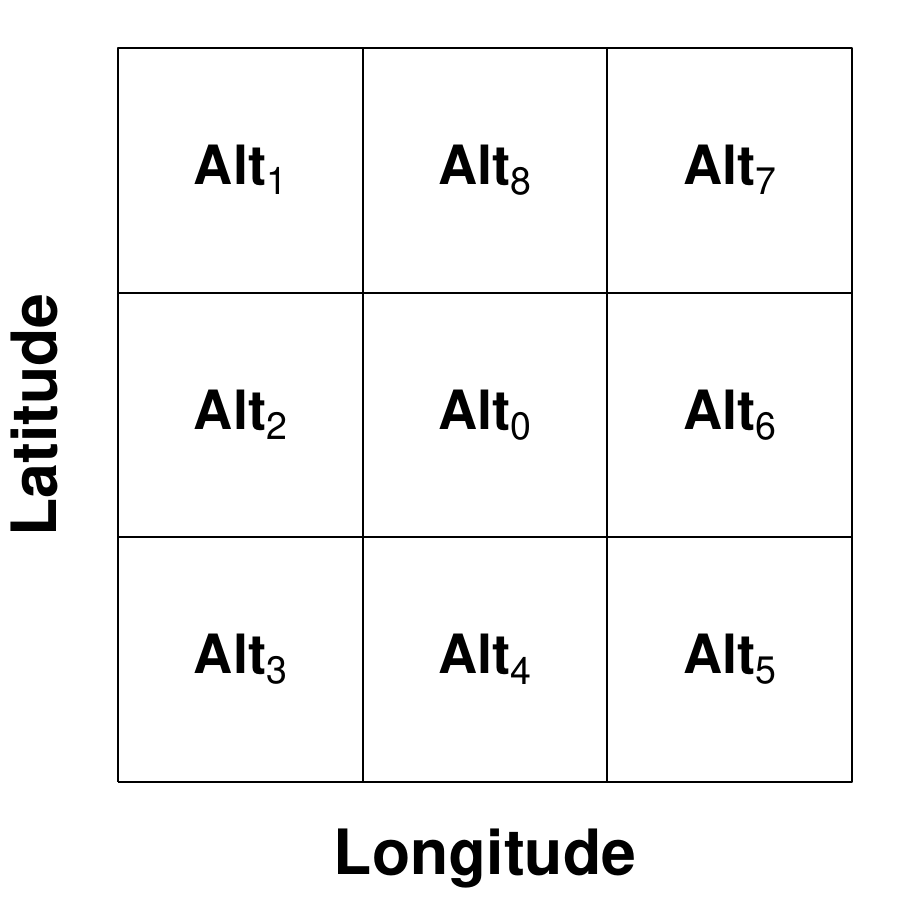}
    \caption{Grid map displaying the relationship between the raster cells used to approximate the variations in the axis directions as used in Equation~\eqref{eq:finite-order}.}
    \label{fig:hornMatrix}
\end{figure}

First, we downloaded the Digital Elevation Model (DEM) data for the Sicily region from the webpage\footnote{URL: \url{https://tinitaly.pi.ingv.it/Download_Area1_1.html}. Last access: Nov. 2023.} of the National Institute of Geophysics and Vulcanology \citep{INGV}. Starting from the DEM of the whole country, we selected the tiles belonging to the Sicily region and combined them to obtain a unique raster object with 10m resolution that represents the altitude in the area of interest. Then, we used the Horn's formula \citep{horn1981hill, weih2004modeling} through the GDAL DEM utility command-line tool \citep{GDAL} to derive the slope from the DEM data. More precisely, the Horn's formula says that the slope $\phi$ can be derived as 
\begin{equation}
\phi = \text{arctan}\left(\sqrt{\frac{\partial \text{Alt}}{\partial \text{lon}} + \frac{\partial \text{Alt}}{\partial \text{lat}}}\right)
\label{eq:Horn}
\end{equation}
where $\text{Alt}$ denotes the altitude dimension and $\partial \text{Alt} / \partial \text{lon}$ and $\partial\text{Alt} /\partial \text{lat}$ represent variations in the East-West and North-South axis, respectively. The two partial derivatives in Equation~\eqref{eq:Horn} can be approximated by using finite-order methods such as 
\begin{equation}
\frac{\partial \text{Alt}}{\partial \text{lon}} = \frac{(\text{Alt}_1 + 2\text{Alt}_2 + \text{Alt}_3) - (\text{Alt}_7 - 2\text{Alt}_6 + \text{Alt}_5)}{8\Delta \text{lon}}
\label{eq:finite-order}
\end{equation}
where $\text{Alt}_1, \dots, \text{Alt}_7$ denote the altitude values in a $3 \times 3$ grid surrounding the cell under analysis (i.e. the cell in the middle of the grid), following the scheme detailed in Figure~\ref{fig:hornMatrix} and $\Delta \text{lon}$ represents the spacing between points in the horizontal direction. Similar considerations hold for the estimation of $\partial \text{Alt} / \partial \text{lat}$ and we refer to \citet{weih2004modeling} and the GDAL DEM webpage\footnote{URL: \url{https://gdal.org/programs/gdaldem.html}} for more details. 

\begin{figure}[!h]
\centering
\includegraphics[width=.45\textwidth, trim = {0 2.5cm 0 2.5cm}, clip]{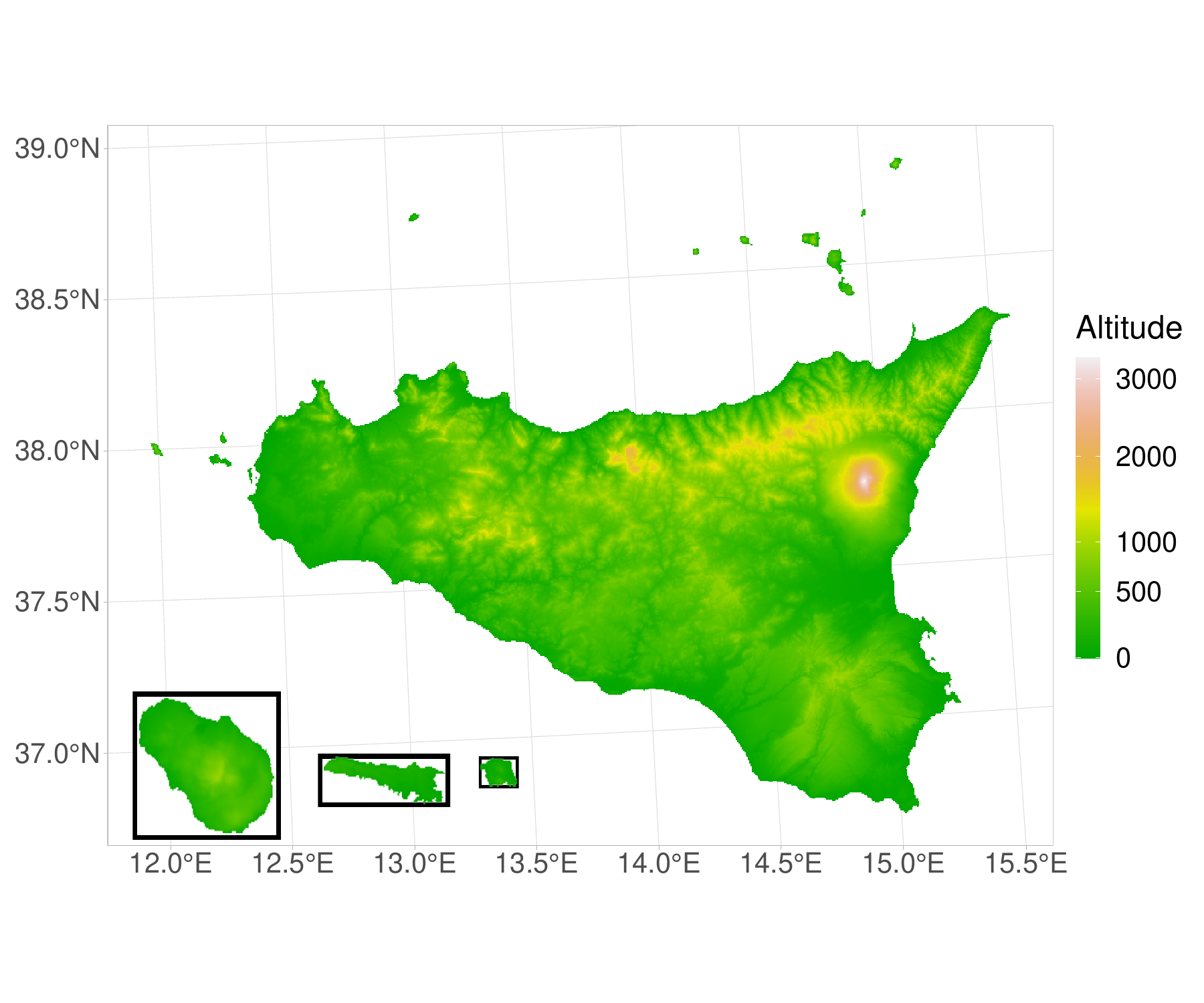}	
\includegraphics[width=.45\textwidth, trim = {0 2.5cm 0 2.5cm}, clip]{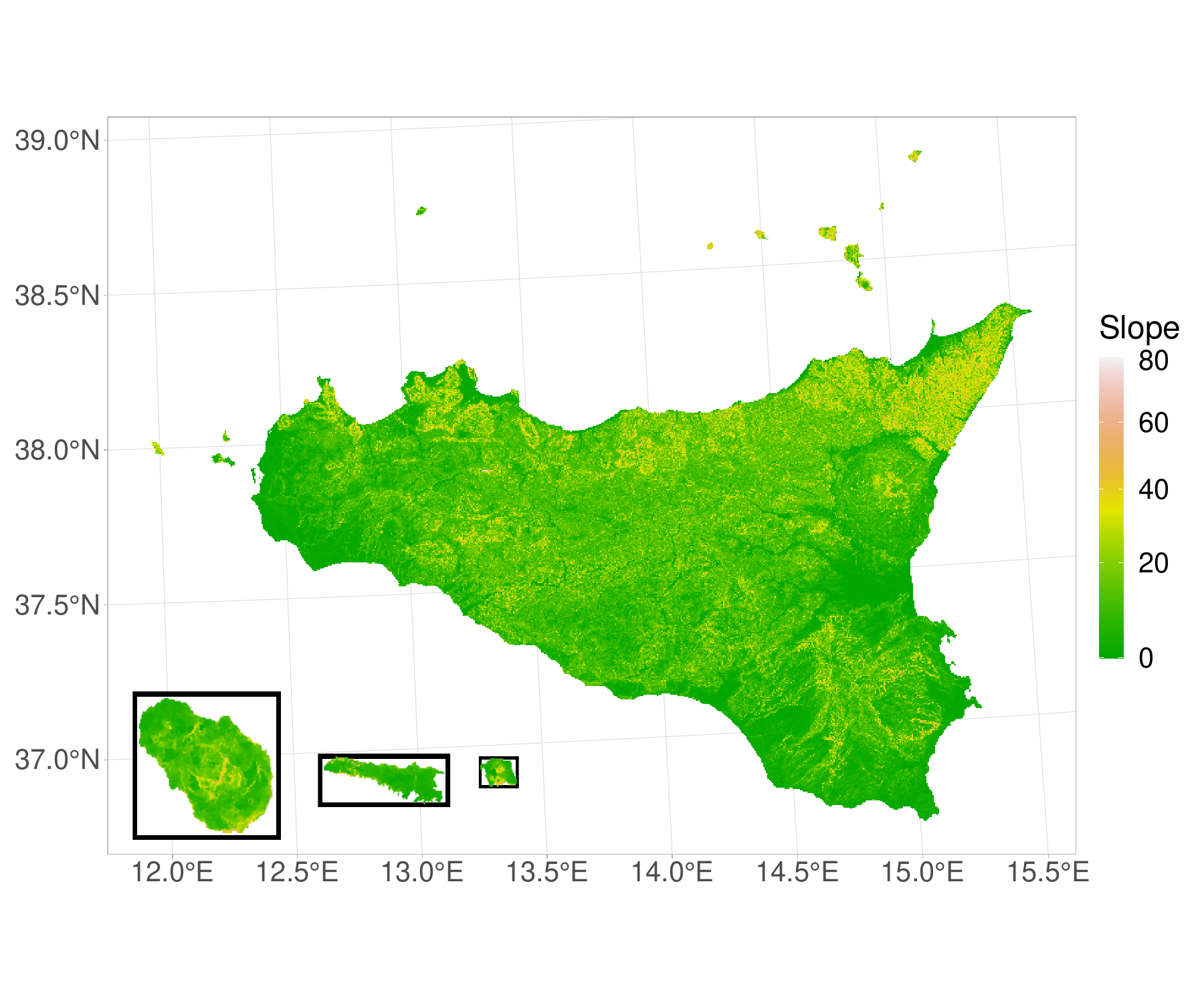}\\
 \includegraphics[width=\textwidth]{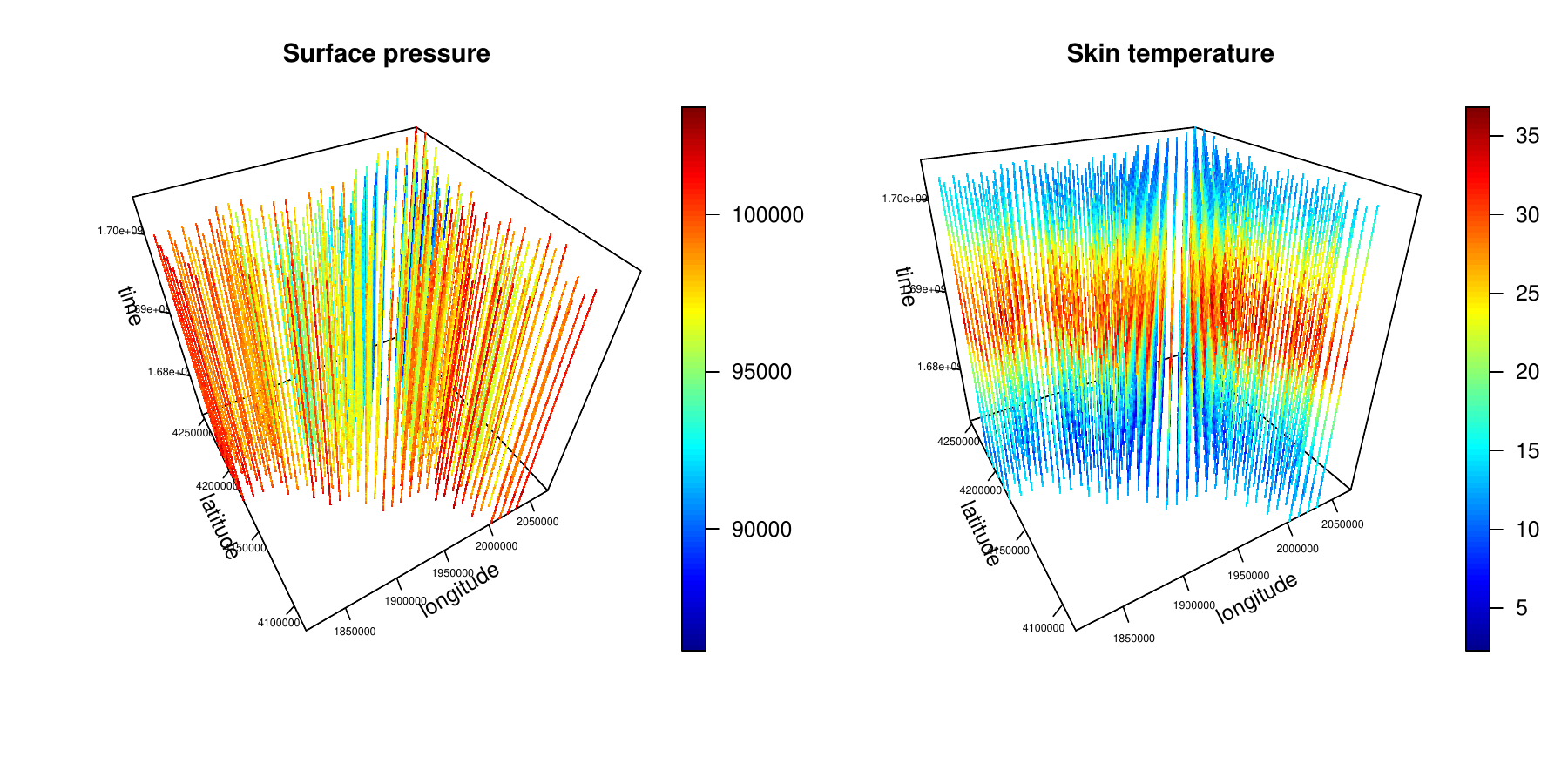}	
	\caption{\textit{Top panels}: Spatial covariates representing the elevation and slope in Sicily: $Z_{elev}(u)$ and $Z_{slope}(u)$;
 \textit{Bottom panels}: Daily means of the surface pressure and skin temperature in Sicily in 2023.}
	\label{fig:4}
\end{figure}

The altitude and slope covariates are displayed in the top panel of Figure~\ref{fig:4}. As we can see, the first map clearly highlights the Etna volcano (located in the eastern part of the region), the mountain chains in the North and North-East areas (Monti Nebrodi), and the Catania valley surrounding the homonymous city. The slope surface is generally flat or almost flat, apart from some regions in the North-East (near Messina).  

Finally, Copernicus ERA5 data contains many environmental covariates that were downloaded from \url{https://cds.climate.copernicus.eu/cdsapp#!/dataset/reanalysis-era5-land?tab=overview}. All of them are obtained in a $38\times34$ spatial grid with $10\text{km} \times 10\text{km}$ raster cells and hourly resolution for each day of the year. The values are averaged daily over the whole year to match the temporal resolution of fires data.
The bottom panel of Figure~\ref{fig:4} contains an example of the spatio-temporal resolution of such variables, and the complete description of the available covariates comes as follows. 

The \texttt{10u} and \texttt{10v} 
variables are the eastward and northward components, respectively, of the 10m wind. They represent the horizontal speed of air moving towards the East and North, at a height of ten metres above the surface of the Earth, in metres per second. In particular, the U wind component is parallel to the x-axis (i.e. longitude) while the V wind component is parallel to the y-axis (i.e. latitude). In particular, a positive U wind comes from the West, and a positive V wind comes from the South.

The surface pressure \texttt{sp} is the pressure (force per unit area) of the atmosphere on the surface of land, sea and in-land water.
It is a measure of the weight of all the air in a column vertically above the area of the Earth's surface represented at a fixed point.
The units of this variable are Pascals (Pa).

The total precipitation \texttt{tp} is the accumulated liquid and frozen water, comprising rain and snow, that falls to the Earth's surface, not including fog, dew or the precipitation that evaporates in the atmosphere before it lands at the surface of the Earth.
This variable is the total amount of water accumulated over a particular time period, which depends on the data extracted. The units of this variable are depth in metres of water equivalent. 

The dew point temperature \texttt{d2m} is the temperature to which the air, at 2 metres above the surface of the Earth, would have to be cooled for saturation to occur. 
The air temperature at 2m above the surface of land, sea or in-land waters is indicated by \texttt{t2m}.
The skin temperature \texttt{skt} is the temperature of the surface of the Earth, that is, the theoretical temperature that is required to satisfy the surface energy balance. It represents the temperature of the uppermost surface layer, which has no heat capacity and so can respond instantaneously to changes in surface fluxes. 
All the soil temperatures \texttt{stl1}, \texttt{stl2}, \texttt{stl3}, and \texttt{stl4}  are the temperature of the soil in the middle of that given layer.
The Advancing global NWP through international collaboration (ECMWF)  Integrated Forecasting System (IFS) has a four-layer representation of soil, where the surface is at 0cm. Layer 1: 0 - 7cm;  Layer 2: 7 - 28cm; Layer 3: 28 - 100cm; Layer 4: 100 - 289cm.
Soil temperature is set at the middle of each layer, and heat transfer is calculated at the interfaces between them. It is assumed that there is no heat transfer out of the bottom of the lowest layer.
All the variables representing temperatures have units of kelvin (K), but have been converted to degrees Celsius (C) by subtracting 273.15 in this paper. They are calculated by interpolating between the lowest model level and the Earth's surface, taking account of the atmospheric conditions.

\section{Spatio-temporal point processes}\label{sec:met}

We consider a spatio-temporal point process with no multiple points as a random countable subset $X$ of $\mathbb{R}^2 \times \mathbb{R}$, where a point $(u, t) \in X$ corresponds to an event at $ u \in \mathbb{R}^2$ occurring at time $t \in \mathbb{R}$.
A typical realisation of a spatio-temporal point process $X$ on $\mathbb{R}^2 \times \mathbb{R}$ is a finite set $\{(u_i, t_i)\}^n_{
i=1}$ of distinct points within a
bounded spatio-temporal region $W \times T \subset \mathbb{R}^2 \times \mathbb{R}$, with area $\vert W\vert  > 0$ and length $\vert T\vert  > 0$, where $n \geq 0$ is not fixed in
advance. 
In this context, $N(A \times B)$ denotes the number of points of a set $(A \times B) \cap X$, where $A \subseteq W$ and $B \subseteq T$. As usual \citep{daley:vere-jones:08}, when $N(W \times T) < \infty $ with probability 1, which holds e.g. if $X$ is defined on a bounded set, we call $X$ a finite spatio-temporal point process.

For a given event $(u, t)$, the events that are close to $(u, t)$ in both space and time, for each spatial distance $r$ and time lag $h$, are given by the corresponding spatio-temporal cylindrical neighbourhood of the event $(u, t)$, which can be expressed by the Cartesian product as
\[
b((u, t), r, h) = \{(v, s) : \vert \vert u - v\vert \vert \leq r, \vert t - s \vert \leq h\} , \quad \quad
(u, t), (v, s) \in W \times T,
\]
where $ \vert \vert \cdot \vert \vert$ denotes the Euclidean distance in $\mathbb{R}^2$ and $  \vert \cdot \vert $ is the absolute value. Note that $b((u, t), r, h)$ is a cylinder with centre $(u, t)$, radius $r$, and height $2h$.

Product densities $\lambda^{(k)},\, k  \in \mathbb{N} \text{ and }  k  \geq 1 $, arguably the main tools in the statistical analysis of point processes, may be defined through the so-called Campbell Theorem (see \cite{daley:vere-jones:08}),  that constitutes an essential result in spatio-temporal point process theory. It states that, given a spatio-temporal point process $X$, for any non-negative function $f$ on $( \mathbb{R}^2 \times \mathbb{R} )^k$

\begin{equation*}
  \mathbb{E} \Bigg[ \sum_{\zeta_1,\dots,\zeta_k \in X}^{\ne} f( \zeta_1,\dots,\zeta_k)\Bigg]=\int_{\mathbb{R}^2 \times \mathbb{R}} \dots \int_{\mathbb{R}^2 \times \mathbb{R}} f(\zeta_1,\dots,\zeta_k) \lambda^{(k)} (\zeta_1,\dots,\zeta_k) \prod_{i=1}^{k}\text{d}\zeta_i,
\label{eq:campbell0}  
\end{equation*}
where $\neq$ indicates that the sum is over distinct values. In particular, for $k=1$ these functions are called the \textit{intensity function} $\lambda$.
Broadly speaking, the intensity function describes the rate at which the events occur in the given spatio-temporal region, representing the point process analogues of the mean function of a real-valued process.
Then, the first-order intensity function is formally defined as 
\begin{equation*}
 \lambda(u,t)=\lim_{\vert \text{d}u \times \text{d}t\vert  \rightarrow 0} \frac{\mathbb{E}[N(\text{d}u \times \text{d}t )]}{\vert \text{d}u \times \text{d}t\vert }, 
\end{equation*}
where $\text{d}u \times \text{d}t $ defines a small region around the point $(u,t)$ and $\vert \text{d}u \times \text{d}t\vert $ is its volume.

\subsection{Spatio-temporal Poisson point processes with separable intensity}\label{sec:method}

The description of the observed point pattern intensity is a crucial issue when dealing with spatio-temporal point pattern data, and specifying a statistical model is a very effective way compared to analyzing data by calculating summary statistics. Formulating and adapting a statistical model to the data allows taking into account effects that otherwise could introduce distortion in the analysis \citep{baddeley2015spatial}.

When dealing with intensity estimation for spatio-temporal point processes, it is quite common to assume that the intensity function $\lambda(u,t)$ is separable \citep{diggle2013statistical,gabriel2009second}. Under this assumption,   the intensity function is given by the product 
 \begin{equation}
   \lambda(u,t)={\lambda}(u){\lambda}(t)
   \label{eq:sep}
 \end{equation}
where ${\lambda}(u)$ and ${\lambda}(t)$ are non-negative functions on $W$ and $T$, respectively.
 Suitable estimates of $\lambda(u)$ and $\lambda(t)$ in Equation \eqref{eq:sep}    depend on the characteristics of each application. This formulation can include a combination of a parametric spatial point pattern model, potentially depending on the spatial coordinates and/or spatial covariates, and a parametric log-linear model for the temporal component. Also, non-parametric kernel estimate forms are legit.
 The spatio-temporal intensity is therefore obtained by multiplying the purely spatial and purely temporal intensities, previously fitted separately. The resulting intensity is normalised, to make the estimator unbiased, making the expected number of points
\[
\mathbb{E}\bigg[ \int_{W \times T}  \hat{\lambda}(u,t)d_2(u,t) \bigg] = \int_{W \times T} \lambda(u,t)d_2(u,t)=n,
\]
and the final intensity function is obtained as
\[
\hat{\lambda}(u,t)=\frac{\hat{\lambda}(u)\hat{\lambda}(t)}{\int_{W \times T} \hat{\lambda}(u,t)d_2(u,t)}.
\]

\subsection{Spatial component}
A general spatial log-linear Poisson model  \citep{cox1972statistical}, generalizes both homogeneous and inhomogeneous models, such that:
\begin{equation}
	\label{eq:loglin}
\lambda(u)= \exp(\boldsymbol{\theta}^{\top} \textbf{Z}(u)),
\end{equation}
where $u \in W $, and $\textbf{Z}(u)=\{Z_1(u), \ldots, Z_p(u)\}$ are known covariate functions, $\boldsymbol{\theta}^{\top}=(\theta_1, \ldots, \theta_p)$ are unknown  parameters, and therefore $\boldsymbol{\theta}^{\top} \textbf{Z}(u)=(\theta_1 Z_1(u)+ \ldots + \theta_p Z_p(u))$.
These models have an especially convenient structure, since the log intensity
is a linear function of the parameters and  covariates can be quite general
functions, making them a very wide class of models.
The estimation of the point process  parameters  is carried out through the maximization of the log-likelihood, defined by:
\begin{equation}
	\label{eq:lik}
	\text{log} \textrm{L}(\boldsymbol{\theta}) = \sum_i \text{log} \lambda(x_i; \boldsymbol{\theta}) -\int_W\lambda(u; \boldsymbol{\theta})du 
\end{equation}
where the sum is over all points $x_i$ in the point process $\textbf{x}$  \citep{daley:vere-jones:08}. For estimation purposes, we use a finite quadrature approximation  of the log-likelihood, 
following \cite{berman1992approximating},
that is, starting from the observed points $u_1,\dots , u_n$, generate $m$  additional \textit{dummy points} ${u}_{n+1} \dots , {u}_{m+n}$ to
form a set of $n + m$ quadrature points (where $m > n$). Then, integral in \eqref{eq:lik} is approximated by the Riemann sum  

$$
    \int_W \lambda({u};\boldsymbol{\theta})\text{d}u \approx \sum_{k = 1}^{n + m}a_k\lambda({u}_{k};\boldsymbol{\theta})
$$
where $a_1, \dots , a_m$ are quadrature weights  and $\sum_{k = 1}^{n + m}a_k = l(W)$, with  $l(\cdot)$ the Lebesgue measure. In order to distinguish between dummy points and observed points, we introduce the indicator $e_k$, which assumes values equal to 1 if the related point is a data point, zero otherwise. Then, writing $y_k = e_k/a_k$, the log-likelihood \eqref{eq:lik} of the template model can be approximated by
$
    \log L(\theta) \approx
\sum_j
a_k
(y_k \log \lambda({u}_k; \boldsymbol{\theta}) - \lambda({u}_k; \boldsymbol{\theta}))
+
\sum_k
a_k.
$
Apart from the constant $\sum_k a_k$, this expression is formally equivalent to the weighted log-likelihood of
a Poisson regression model with responses $y_k$ and means $\lambda({u}_k,t_k; \theta) = \exp(\theta Z(u_k,t_k))$. 
This means that the model can be
maximised using standard GLM software, but also that covariate values must be known in every data and dummy point location. Indeed, the spatial covariates are referred to as those variables with observable values, at least in principle, at each spatial location in the spatial window, and for the inferential purposes given above, their values must be known at each point of the data point pattern and at least at some other locations. 

\subsubsection{Diagnostics}\label{sec:diag}

For an inhomogeneous Poisson process model, with fitted intensity $\hat{\lambda}(u)$, the predicted number of points falling in any region $W$ is $\int_W \hat{\lambda}(u)$d$u$. Hence, the residual in each region $W \subset \mathbb{R}^2$ is  the `observed minus predicted' number of points falling in $W$ \citep{alm1998approximation}, that is $ R(W)=n(\textbf{x} \cap W) - \int_W \hat{\lambda}(u)du$, where \textbf{x} is the observed point pattern, $n(\textbf{x} \cap W)$ the number of points of \textbf{x} in the region $W$, and $  \hat{\lambda}(u)$ is the intensity of the fitted model.
A simple residual visualization can be obtained by smoothing them.
The `smoothed residual fields' are defined as
\begin{equation}
s(u)=\tilde{\lambda}(u)-\lambda^{\dagger}(u)
\label{eq:smo0}
\end{equation}
where $\tilde{\lambda}(u)=e(u)\sum_{i=1}^{n(\textbf{x})} \kappa(u-\textbf{x}_i)
\label{eq:smo1}$  is the non-parametric, kernel estimate of the fitted intensity $ \hat{\lambda}(u)$, while $\lambda^{\dagger}(u)$ is a correspondingly-smoothed version of the (typically parametric) estimate of the intensity of the fitted model, $ \lambda^{\dagger}(u)=e(u) \int_W \kappa(u-v)\hat{\lambda}(v)dv$. Here, $\kappa$ is the smoothing kernel and $e(u)$ is the edge correction. 
The smoothing bandwidth for the kernel estimation of the raw residuals is selected by cross-validation,  as the value that minimises the Mean Squared Error criterion defined by \cite{diggle1985kernel} by the method of \cite{berman1989estimating}. 

The differences in Equation \eqref{eq:smo0} should be approximately zero when the fitted model is close to the real one.
Therefore, the best model is the one with the lowest values of the smoothed raw residuals. Their graphical representation gives further insights on which spatial regions the model could not be a good fit.

\subsection{Temporal component}

Temporal count data analysis often involves modelling the occurrence of events over time, and a suitable statistical approach is essential to capture the temporal dynamics adequately. In this context, the Poisson   regression model, which belongs to the class of Generalized Linear Models (GLMs), provides a robust framework for analyzing count data exhibiting temporal variation.

Consider a set of covariates $  \boldsymbol{Z}(t) = \{ Z_1(t),\ldots, Z_q(t)  \}$ influencing the event rate. The temporal count Poisson model is then formulated as follows:
\[
\lambda(t) = \exp(\boldsymbol{\beta}^T \boldsymbol{Z}(t)).
\]

Here, $t \in T$ and $\boldsymbol{\beta} =  \{\beta_0, \beta_1,  \ldots \beta_q\}$ are the model parameters associated with the temporal covariates.

The model parameters are estimated through the maximum likelihood estimation (MLE) method within the GLM framework \citep{GLM}. 

After fitting the model, assessing its performance involves standard GLM techniques such as residual analysis, goodness-of-fit tests, and validation on independent datasets. Interpretation of the model parameters provides insights into the impact of covariates and temporal components on the event rate over time.

A generalized additive model (GAM) is a GLM in which the linear predictor is given by a sum of smooth functions of the covariates plus a conventional parametric component. In such a case, we may write
$
\lambda(t) = \exp(f(Z(t)))
$
where 
$f(\cdot)$
  is a smooth function of the covariate 
$Z$. 
The smooth terms can be functions of any number of covariates, and the control over the smoothness of the functions is solved by using the Generalized Cross Validation (GCV) criterion \citep{wood2017generalized}.

\section{Results}\label{sec:int}

In this section, we present the results obtained fitting the separable spatio-temporal intensity function presented in Section~\ref{sec:met} to the data introduced in Section~\ref{sec:analysis}. Specifically, in Section \ref{sec:spazio}, a spatial log-linear Poisson point process model is applied to investigate the influence of land use types on the purely spatial fire distribution, accounting for other environmental covariates. In Section \ref{sec:tempo}, a Poisson Generalized Additive Model is fitted to the temporal fire occurrences, depending on both non-parametric components of the temporal coordinates and parametric formulation of environmental covariates.
 
\subsection{Spatial intensity}\label{sec:spazio}

Starting from the maximal model and proceeding with a backwards procedure based on the comparison of the AIC values and the sequential performance of tests for the significance of the model parameters, the chosen spatial model has a linear predictor that includes a non-parametric term for the spatial coordinates and parametric expression for the spatial covariates, as follows
\begin{equation}
	\label{eq:space}
\lambda(u)=\exp(f(u)+ \boldsymbol{\theta}_1Z_{land\_use}(u)+\theta_2Z_{elev}(u)+\theta_3Z_{slope}(u)).
\end{equation} 
Here, $f(\cdot)$ is a nonparametric function for $u \in W$, estimated through thin plate regression splines \citep{wood2003thin} with 30 knots.

Table \ref{tab:regression_results} presents \add{the} results from the fitted spatial Poisson model that investigates the influence of various factors on fire occurrence. In addition to the intercept, the first three rows correspond to the categories of the \textit{land use} variable, with \textit{Artificial surfaces} serving as the baseline. Also note that the elevation has been converted from meters to kilometres, to obtain estimated coefficients on the same magnitude as the other ones. 

\begin{table}
  \centering
\begin{tabular}{lrrrr}
    \toprule
 & \textbf{Estimate} & \textbf{Std. Error} & \textbf{z value} & \textbf{p value}  \\
    \midrule
    \textbf{Intercept} & -15.129 & 0.040 & -376.580 & $< 1 \times 10^{-3}$ \\
    \textbf{Agricultural areas}  & -0.286 & 0.040 & -7.179 & $< 1 \times 10^{-3}$ \\
    \textbf{Forest and semi-natural areas}  & -0.540 & 0.046 & -11.727 & $< 1 \times 10^{-3}$ \\
    \textbf{Wetlands and water bodies} & -0.595 & 0.154 & -3.870 & $< 1 \times 10^{-3}$ \\
    \textbf{Elevation} & 1.675 & 0.040 & 42.320 & $< 1 \times 10^{-3}$ \\
    \textbf{Slope} & 0.006 & 0.001 & 4.768 & $< 1 \times 10^{-3}$ \\
    \bottomrule
\end{tabular}  
\caption{Estimated coefficients of the fitted spatial model in Equation \eqref{eq:space}.}
  \label{tab:regression_results}
\end{table}

Considering different land usages, agricultural areas show a significant negative estimated parameter, signifying a considerable decrease in the fire occurrence intensity moving from artificial surfaces to agricultural ones.
Forest and semi-natural areas also show a negative effect, although with a higher magnitude compared to agricultural areas.
Then, wetlands and water bodies also display a significant negative parameter, albeit with a wider confidence interval, indicating greater uncertainty. This is likely due to the sparsity of the area covered in Sicily by this type of land.
Lastly, both elevation and slope exhibit a significant effect, with their increase being associated with higher intensity.
Note that an attempt has been made to assess the significance of the ERA5 spatio-temporal covariates, averaged over time. However, these purely spatial covariates did not influence the overall spatial occurrence of fires and therefore were not included in the chosen model. This is likely due to the very coarse grid of the obtained spatial covariates if compared to those representing elevation and slope (top panels of Figure \ref{fig:4}).

\begin{figure}[!h]
\centering	\includegraphics[width=.45\textwidth]{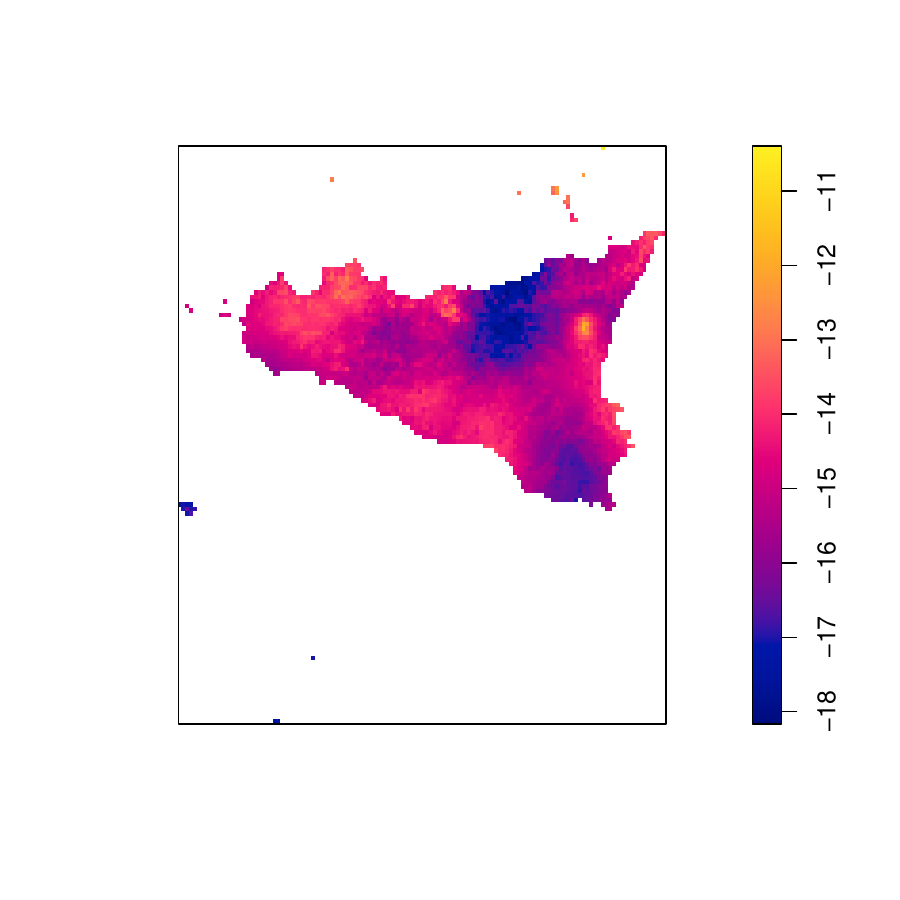}
 \includegraphics[width=.45\textwidth]{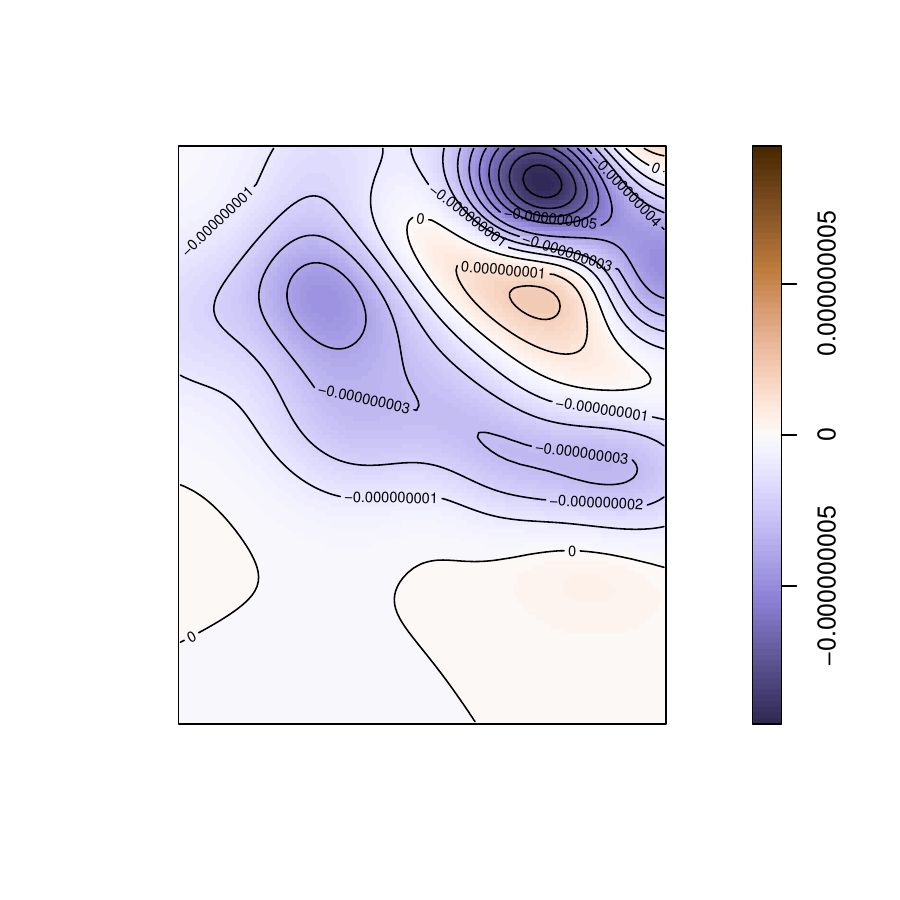}
	\caption{\textit{Left panel}: Intensity predicted according to the fitted spatial model in Equation \eqref{eq:space}; \textit{Right panel}: Smoothed raw residuals of the fitted model.}
	\label{fig:space_predict}
\end{figure}

Among the perks of fitting a semi-parametric model as the one in Equation \eqref{eq:space} there is the smoothed prediction map that can be obtained, as shown in the left panel of Figure \ref{fig:space_predict}. The smoothed raw residuals in the right panel of Figure \ref{fig:space_predict}, computed as detailed in Section \ref{sec:diag}, indicate a good model fitting, exhibiting almost all residuals approaching zero.

\subsection{Temporal intensity}\label{sec:tempo}

Following the same backward procedure adopted for the spatial component, the chosen temporal model comes with a linear predictor that includes a non-parametric term for temporal coordinates and parametric expression for the temporal covariates, as follows:
\begin{equation}
	\label{eq:time}
\lambda(t)=\exp(f(t)+\beta_1Z_{v10}(t)+\beta_2Z_{stl2}(t)+\beta_3Z_{sp}(t)+\beta_4Z_{tp}(t)),
\end{equation} 
with $f(\cdot)$ a nonparametric function for $t \in  T$, estimated through penalized regression basis splines \citep{wood2017generalized} with 50 knots.

\begin{table}[ht]
\centering
\begin{tabular}{lrrrr}
\toprule
 & \textbf{Estimate} & \textbf{Std. Error} & \textbf{z value} & \textbf{p value} \\
\midrule
\textbf{Intercept} & 77.314 & 5.471 & 14.133 & $< 1 \times 10^{-3}$ \\
\textbf{Wind speed from South} & 0.232 & 0.008 & 30.426 & $< 1 \times 10^{-3}$ \\
\textbf{Temperature} & 0.440 & 0.024 & 18.672 & $< 1 \times 10^{-3}$ \\
\textbf{Surface pressure} & -0.001 & 0.000 & -16.251 & $< 1 \times 10^{-3}$ \\
\textbf{Total precipitation} & -14.848 & 2.266 & -6.553 & $< 1 \times 10^{-3}$ \\
\bottomrule
\end{tabular}
\caption{Estimated coefficients of the fitted temporal model in Equation \eqref{eq:time}.}
\label{tab:temp1}
\end{table}

  \begin{figure}[!h]
  \centering
  \includegraphics[width=.495\textwidth]{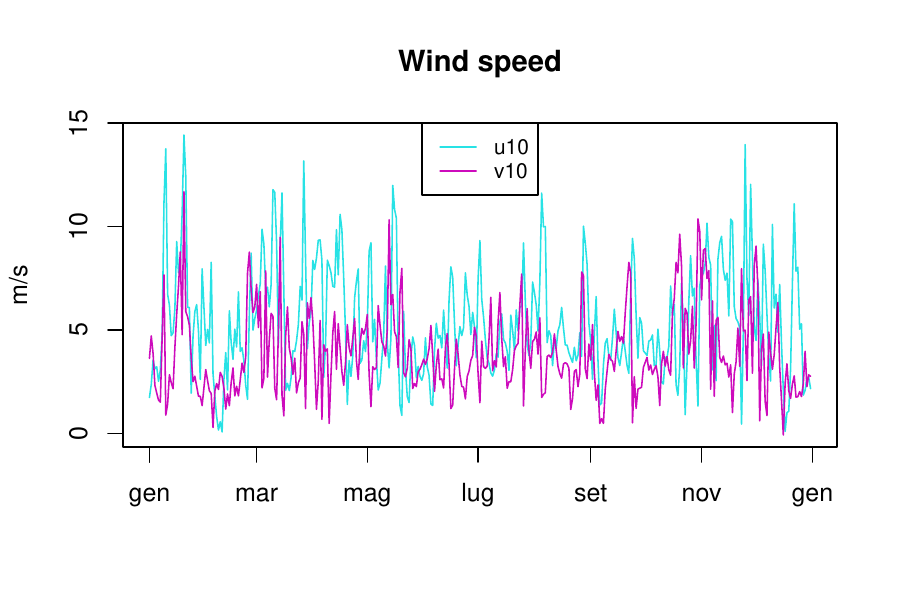}
\includegraphics[width=.495\textwidth]{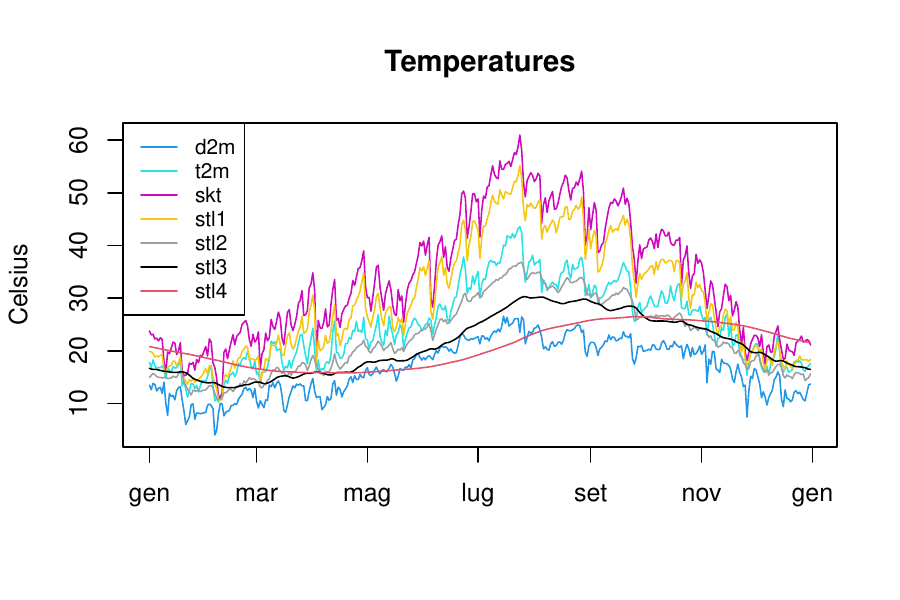}	\\
 \includegraphics[width=.495\textwidth]{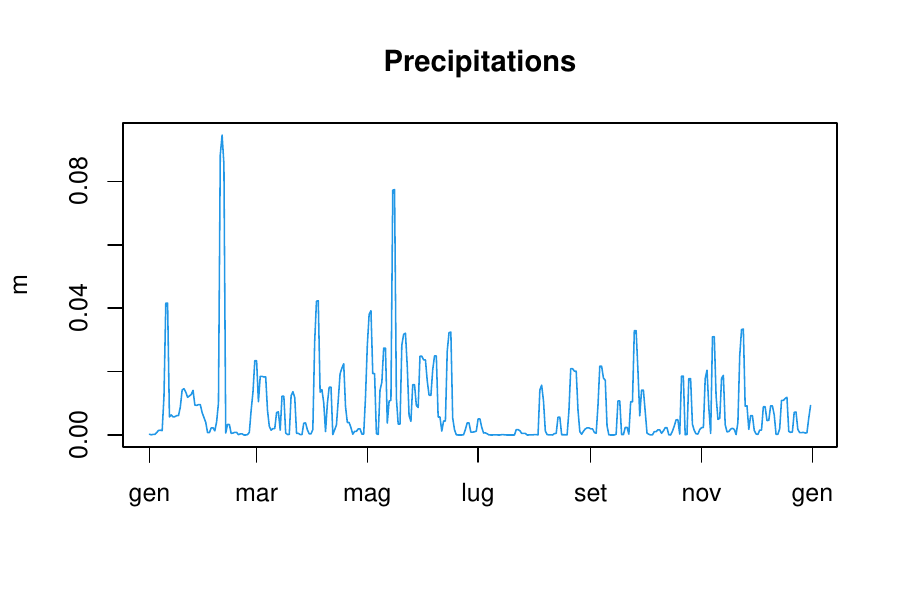}	
 \includegraphics[width=.495\textwidth]{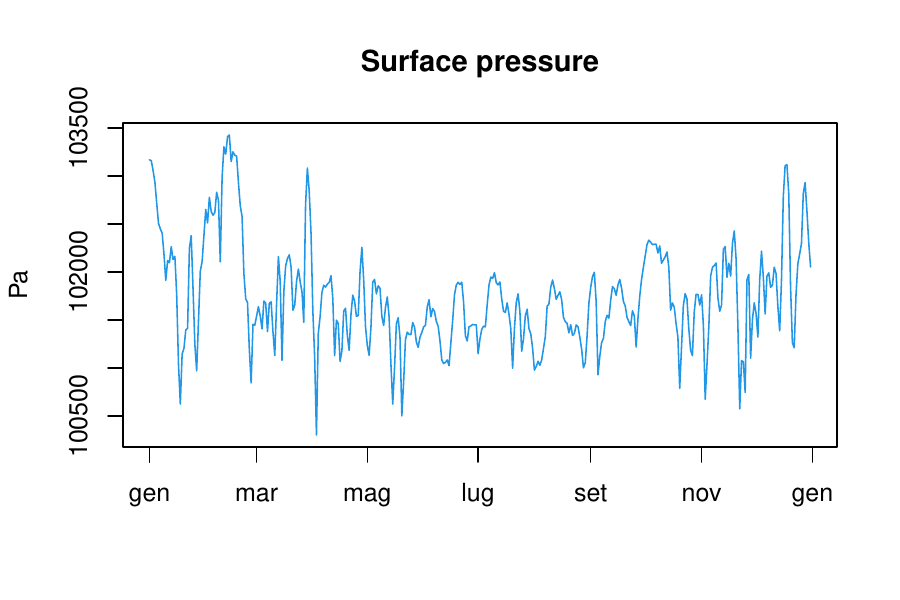}
	\caption{Daily maxima of the wind components, temperatures, precipitation and surface pressure}
	\label{fig:4time}
\end{figure}

\begin{figure}[!h]
  \centering
 	\includegraphics[width=\textwidth]{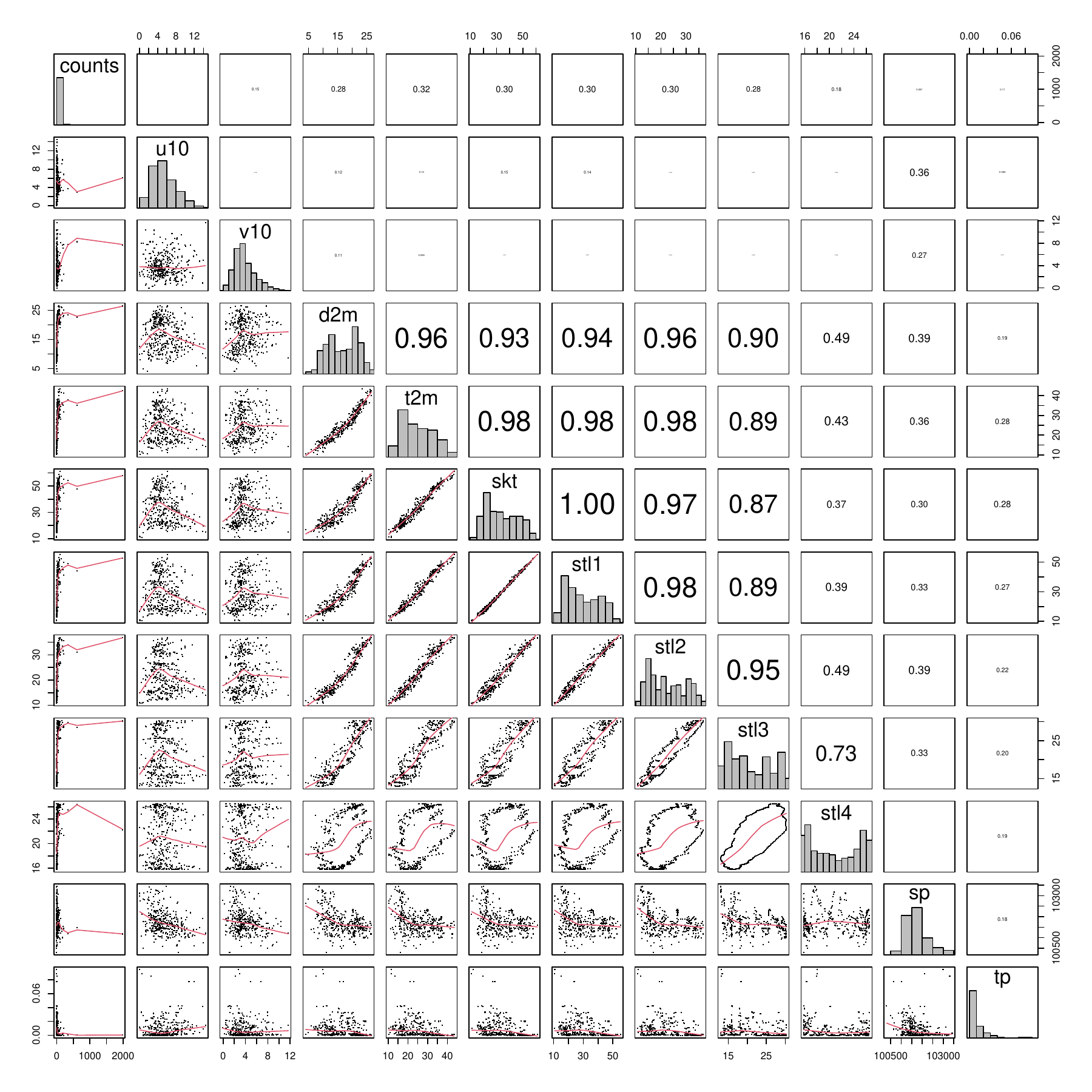}	
	\caption{Correlations among the temporal covariates and the daily fire counts over 2023. On the main diagonal, the univariate distributions of the variables. On the lower panels, the scatterplots of each pair of variables with a smoothing function overlapped in red. On the upper panels, the correlations among the variables.}
	\label{fig:4b}
\end{figure}

In order to fit this model, the ERA5 environmental spatio-temporal covariates have been averaged with respect to the spatial components, making them purely temporal covariates. Moreover, the results of the models fitted with the daily means of such covariates have been compared to those containing the daily maxima, and the latter yielded better fitting. The resulting temporal covariates are shown in Figure \ref{fig:4time}. As evident, the temperatures appear highly correlated among themselves and with the time of the year. In detail, the deeper the surface layers of detection, the smoother the temperature variability.  Figure \ref{fig:4b} further illustrates such high correlations, especially among temperature variables. 
Table \ref{tab:temp1} outlines the outcomes of the fitted model, whose detailed breakdown follows. 

Wind speed demonstrates a notable positive influence, implying that higher wind speeds from the South correlate with increased fire counts, highlighting the role of the Scirocco wind in wildfire dynamics. This Mediterranean wind that comes from the Sahara brings heat and dust from African coastal regions.
Temperature shows a significant positive relationship with fire counts.
Pressure exhibits a negative estimated parameter, suggesting that higher atmospheric pressure is associated with lower fire counts, hinting at its potential suppressive effect.
Precipitation reveals a substantial negative effect, indicating fewer fire counts with higher daily precipitation, highlighting its mitigating effect.

The significant smooth terms suggest that time plays a crucial role in the observed residual variability of fire incidents, not taken into account by the environmental covariates. 

\begin{figure}[!h]
  \centering
 	\includegraphics[width=.75\textwidth]{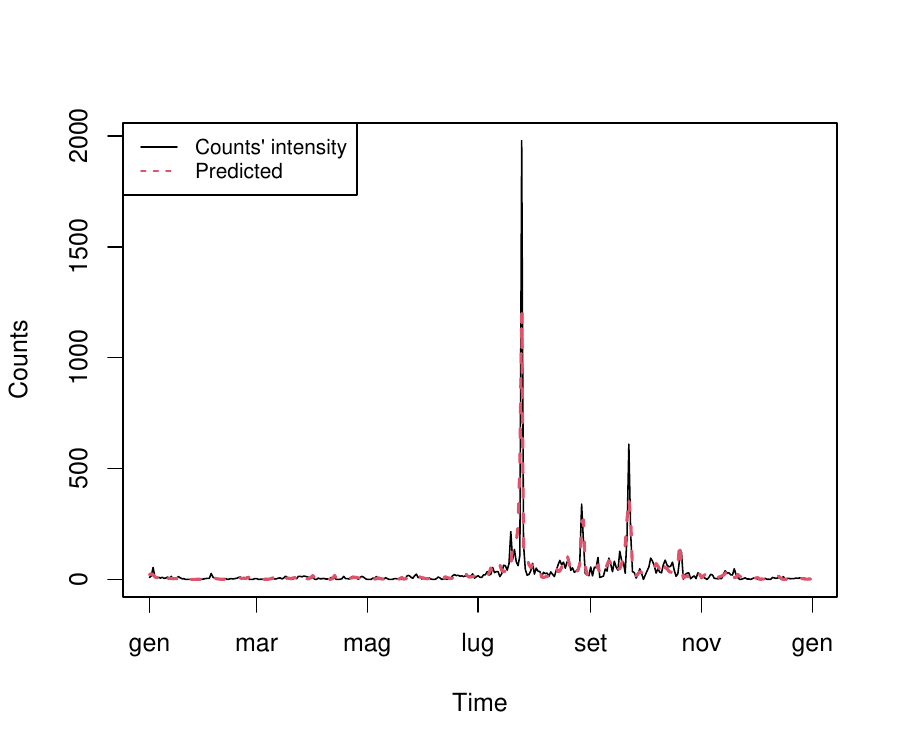}
	\caption{Temporal intensity predicted according to the fitted temporal model in Equation \eqref{eq:time} (red dashed line) and the observed fire counts (solid line).}
	\label{fig:imp}
\end{figure}

An adjusted R-squared of 0.738 and an 83.9\% of deviance explained indicates that the model explains substantial variability in fire counts.
Figure \ref{fig:imp} shows the intensity predicted according to the fitted model in Equation \eqref{eq:time}, demonstrating a good ability to capture the temporal variability of fire counts.

\section{Discussion and conclusions}\label{sec:concls}

The proposed analysis of the spatial and temporal distribution of fire occurrences in Sicily during 2023 has provided insights into the factors influencing these events, with a particular emphasis on the role of land usage. Our findings highlight the significance of human activities in contributing to the prevalence of fires.
The identification of artificial surfaces as a key contributor to the increased probability of fire occurrence highlights the urgent need for targeted intervention and policy measures. The implications of this research are particularly pertinent, considering the well-established fact that the majority of fires in Sicily are human-induced \citep{ferrara2019background}. 
These results have broader implications for regional planning, resource allocation, and the development of proactive measures to manage fire risks effectively. Recognizing the human-centric nature of fire occurrences prompts us to advocate for sustainable land-use practices and the implementation of policies aimed at minimizing the risk of fires in areas characterized by artificial surfaces.
Our analysis extends beyond land usage, incorporating environmental covariates that significantly influence the occurrence of fires in Sicily during 2023. Our findings reveal a significant role of these environmental factors over the temporal event occurrences.
Particularly, wind speed from the South and temperature emerge as crucial variables, emphasising the importance of climatic conditions in amplifying fire risks. 
Conversely, our study identifies surface pressure and total precipitation as environmental covariates that negatively influence fire occurrences as mitigating factors.
Additionally, our research highlights the significance of terrain characteristics in explaining the spatial distribution of fire points. In particular, both elevation and slope exhibit a positive effect on the occurrence of fires, indicating that areas with higher elevations and steeper slopes are more susceptible. 



Note that other FIRMS variables could have been used in this paper. These are the brightness temperature in Kelvin, the along scan and track pixel sizes, the type of satellite, a confidence value attached to each individual hotspot/fire pixel, the collection/version source, the Fire Radiative Power, in megawatts, and whether the fire occurred during daytime or nighttime.
The main reason for not using this type of covariates is that, in point process theory, they are referred to as ``marks" of the observed point pattern \citep{daley:vere-jones:08}.
While the spatial covariates are referred to as those variables with observable values, at least in principle, at each spatial location in the spatio-temporal window, marks are characteristics of the events, which are attached to the points of the observed patterns and can be studied to explore their eventual aid in describing the phenomenon under study.
Formally, however, their inclusion in a regression-type model is quite different if compared to the will to include the so-called spatial or spatio-temporal covariates. Indeed, the latter covariates are characteristics of the spatial or spatio-temporal region, and their inclusion in a regression-type point process model is more straightforward, not needing any particular assumption on, for instance, their distribution.         
Moreover, the literature on the so-called marked point process methodologies is rather specific to particular physical phenomena, like the seismic ones with the ETAS models \citep{ogata:88}, and quite limited for the cases in the presence of multiple marks. This is why we decided not to include marks in our analysis, even though we believe it represents a very interesting topic, both from the methodological and the applied points of view.

Moving to other possible future research paths, a major topic that could be addressed is the definition and application of tailored diagnostic procedures to assess the need for more complex models in case of residual clustering behaviour of points not taken into account by the already considered factors. Indeed, the employment of most known diagnostic tools based on second-order summary statistics \citep{adelfio:schoenberg:09,adelfio2020some}, both global and local, becomes computationally hard as the number of points increases.
In a preliminary way, it is worth noticing that one could first run a test for first-order separability prior to fitting the model. In particular, \cite{schoenberg2004testing}, \cite{diaz2013similarity}, and \cite{fuentes2018first} show that the intensity of forest fire occurrences varies in space and time in a nonseparable way. We, however, believe that our proposed model overcomes computational burden issues, typical of more complex non-separable spatio-temporal models.

Moreover, the inclusion and test of spatially (or spatio-temporally) varying covariates in intensity function has been of
particular interest in \cite{diaz2014significance}, \cite{borrajo2017testing,borrajo2020bootstrapping,borrajo2020testing} and \cite{myllymaki2021testing}.
Following these, future work includes the assessment of separability assumption, as well as exploring more complex models, like the log-Gaussian Cox processes, multitype Poisson models, and local ones  \citep{dangelo2023locally}.

\section*{Fundings}
The research work of Nicoletta D'Angelo and Giada Adelfio was supported by:
 \begin{itemize}
     \item Targeted Research Funds 2024 (FFR 2024) of the University of Palermo (Italy);
     \item Mobilità e Formazione Internazionali - Miur INT project ``Sviluppo di metodologie per processi di punto spazio-temporali marcati funzionali per la previsione probabilistica dei terremoti";
     \item European Union -  NextGenerationEU, in the framework of the GRINS -Growing Resilient, INclusive and Sustainable project (GRINS PE00000018 – CUP  C93C22005270001).
 \end{itemize}

Andrea Gilardi acknowledges the support by MUR, grant Dipartimento di Eccellenza 2023-2027. His research work is funded by the European Union - NextGenerationEU, in the framework of the GRINS - Growing Resilient, INclusive and Sustainable project (GRINS PE00000018 – CUP D43C22003110001). 

The research work of Alessandro Albano has been supported by the European Union - NextGenerationEU - National Sustainable Mobility Center CN00000023, Italian Ministry of University and Research Decree n. 1033— 17/06/2022, Spoke 2, CUP B73C2200076000 and Targeted Research Funds 2024 (FFR 2024) of the University of Palermo (Italy).

The views and opinions expressed are solely those of the authors and do not necessarily reflect those of the European Union, nor can the European Union be held responsible for them.

\bibliography{biblio}

\end{document}